\documentclass{article}

\usepackage[colorlinks,urlcolor=blue,linkcolor=blue,citecolor=blue]{hyperref}

\usepackage{color,array}
\usepackage{amsmath,amssymb,amsfonts}
\usepackage{algorithm2e}
\usepackage{graphicx}
\usepackage{subcaption}
\usepackage[top=0.75in, bottom=0.79in, left=0.75in, right=0.75in]{geometry}

\newtheorem{theorem}{Theorem}
\newtheorem{lemma}{Lemma}
\newtheorem{proposition}{Proposition}
\begin{document}


\title{\textbf{Tempering the Bayes Filter towards Improved Model-Based Estimation}} 


\author{M. J. T. C. van Zutphen, D. Herceg, G. Delimpaltadakis, D. Antunes}
\date{}





\maketitle

\textbf{\textit{abstract:}}
Model-based filtering is often carried out while subject to an imperfect model, as learning partially-observable stochastic systems remains a challenge. Recent work on Bayesian inference found that tempering the likelihood or full posterior of an imperfect model can improve predictive accuracy, as measured by expected negative log likelihood. In this paper, we develop the tempered Bayes filter, improving estimation performance through both of the aforementioned, and one newly introduced, modalities. The result admits a recursive implementation with a computational complexity no higher than that of the original Bayes filter. Our analysis reveals that --- besides the well-known fact in the field of Bayesian inference that likelihood tempering affects the balance between prior and likelihood --- full-posterior tempering tunes the level of entropy in the final belief distribution. We further find that a region of the tempering space can be understood as interpolating between the Bayes- and MAP filters, recovering these as special cases. Analytical results further establish conditions under which a tempered Bayes filter achieves improved predictive performance. Specializing the results to the linear Gaussian case, we obtain the tempered Kalman filter. In this context, we interpret how the parameters affect the Kalman state estimate and covariance propagation. Empirical results confirm that our method consistently improves predictive accuracy over the Bayes filter baseline.

\section{Introduction} 

Accurate knowledge of system states is crucial to the implementation of effective control strategies~\cite{thrun2005probabilistic}. In many real-world applications like financial markets, meteorology, social media user behavior, and robot localization, system states are only partially-observable. When full state measurements are unavailable, an observer or filtering procedure is required to generate state estimates from output measurements. In practice, model-based filtering applications are often subject to significant model mismatch due to the challenging nature of stochastic dynamics identification from data. Techniques such as the Baum–Welch algorithm~\cite{baum1970maximization} and more recent machine learning identification methods~\cite{mann2023endtoend,rimella2025hidden} have demonstrated that relatively accurate hidden Markov models (HMMs) can often be identified directly from data. However, methodological shortcomings like simplifying distribution assumptions~\cite{Zheng2021LH}, insufficiently rich model parameterizations~\cite{Mevel2004AS}, and/or a limited amount of available training data, generally still means that model imperfections persist.

The classic Bayes-~\cite{thrun2005probabilistic} and maximum a posteriori (MAP)~\cite{bishop2006pattern} filters have long provided fundamental frameworks for the state estimation problem. Both enable the recursive update of estimates with each new observation. The Bayes filter propagates the entire state probability distribution (the belief) over time, whereas MAP estimators propagate only the single most probable state sequence, both rooted in the same Bayesian probabilistic framework. The celebrated Kalman filter~\cite{Klmn1960ANA} is a special case of this framework for linear-Gaussian systems, yielding an optimal closed-form solution for the posterior mean and covariance. All these filters critically rely on the accuracy of the available stochastic system model. Even mild model mismatch or uncertainty can significantly degrade estimator performance in practice \cite{Mevel2004AS}. 

Model-free data-based methods have been proposed to bypass model-based approach issues entirely~\cite{Wolff2024RD}. The resulting MAP-based moving horizon estimator (MHE) method performs comparably to standard model-based MHE, and outperforms it when model bias is present. Unfortunately, highly performant model-free methods can be computationally expensive; the aforementioned model-free MHE approach~\cite{Wolff2024RD}, for example, requires solving a constrained QP at each step.  Alternatively, methods have been developed that protect against --- instead of compensating for --- the issue of model mismatch. For example, the paper~\cite{Ramezani2002HM}, extends standard HMM estimation theory by embedding model and data uncertainty information directly into the filtering mechanism via a risk-sensitive exponential cost, providing a framework for estimation that, although conservative, remains stable and reliable even under moderate model mismatch. 

Recent results in the adjacent field of \emph{Bayesian inference machine learning} have demonstrated that the computationally cheap \emph{distributional tempering} method improves estimation performance for a wide range of data-based classifiers~\cite{Wenzel2020HowGI}. In a large set of relevant examples, tempered posteriors have been shown to consistently outperform their untempered counterparts, despite Bayesian theory predicting optimality of models obtained through the Bayesian inference framework. Here, we outline a short history of these tempering methods in the literature. 

In early PAC-Bayes bounds literature, the likelihood-tempering form naturally appears~\cite[Sec.~1.1]{Catoni2007PACBayesian}. In particular, Catoni’s classic bound is optimized by balancing a data log-likelihood term against a \emph{complexity} term (KL-divergence from the prior), with trade-off factor $1/\lambda_{L}$. The likelihood-tempering parameter $\lambda_{L}$ regulating the influence of the prior versus that of the data model is an interpretation of likelihood tempering that is often invoked in later literature. However, the idea that tuning this balance could improve performance of data-based Bayesian methods was recognized only much later. A paper investigating generalizations of the Bayesian updating rule that was published shortly afterward explicitly concludes that $\lambda_{L}\ne 1$ can not be considered coherent (generalized) Bayes~\cite{Bissiri2013AGF}. Soon thereafter, the work~\cite{Mandt2014VariationalT} experiments with initially increasing $\lambda_{L}$ to encourage exploration (as opposed to exploitation). 
However, their implicit assumption remains that estimation should be done with $\lambda_{L}=1$ once training is complete. 

The \emph{Safe Bayes} algorithm~\cite{grunwald2017inconsistency} is one of the first methods proposing that permanent likelihood tuning can improve Bayesian inference performance. When model-class misspecification causes the likelihood to concentrate around a finite number~${(>1)}$ of incorrect models, the method proposes $\lambda_{L}\in(0,1)$ to flatten the posterior enough to concentrate it around a single coherent compromise region. A number of papers that empirically experiment with likelihood tempering then appear. A heuristic likelihood-tempering (flattening $\lambda_{L}\in(0,1)$) factor is proposed to avoid overconfident and abrupt model-probability changes during online inference~\cite{Reichl2017}. Likelihood tempering is used as a tunable hyperparameter after the authors recognize $\lambda_{L}$ in the evidence lower-bound (ELBO) maximization objective~\cite{Bae2018EigenvalueCN}. As it is interpreted as a somewhat arbitrary hyperparameter, it still receives little explicit attention. A tempered likelihood with $\lambda_{L}\in[1,\infty)$ is further applied in large-scale Bayesian deep learning for better calibration and training stability~\cite{Osawa2019PracticalDL}. Letting $\lambda_{L}$ start large and converge to one yields an initial dampening effect on the gradient descent step.

The seminal paper~\cite{Wenzel2020HowGI} is the first in the Bayesian-inference machine-learning field to remark that tempering of the Bayesian inference \emph{full posterior} (specifically: cooling $\lambda_{P}\in(1,\infty)$) can improve estimator performance. They demonstrate this on a number of examples while noting that this is not consistent with Bayes’ theorem~\cite{Bissiri2013AGF}, forming the start of a research line that investigates this surprising \emph{cold-posterior effect}. While claiming to investigate the cold-posterior effect as described by~\cite{Wenzel2020HowGI}, most papers end up studying the beneficial effects of the adjacent \emph{likelihood tempering} instead. The subsequent paper~\cite{Adlam2020ColdPA}, for instance, provides a theoretical exploration of likelihood tuning $\lambda_{L}\in(1,\infty)$ in its search for possible explanations for the cold-posterior effect. It conjectures the source of the effect could be likelihood misspecification, curation, or incorrect handling of data augmentation. 

A series of works~\cite{Aitchison2020AST,Nabarro2021DataAI,Noci2021DisentanglingTR,Bachmann2022HowTF} further investigate the cold-posterior effect. They note that the likelihood-tempered form $\lambda_{L}>1$ emerges naturally when correcting for the fact that \emph{multiple} humans had to agree on each label in curated datasets~\cite{Aitchison2020AST}, even arguing against full-posterior tempering. They hypothesize that the improper implementation of data augmentation (DA) causes the cold-posterior effect, but show that this hypothesis in not correct through controlled experiments~\cite{Nabarro2021DataAI}. The theoretical follow-up~\cite{Bachmann2022HowTF} provides results explaining how likelihood tempering corrects for the i.i.d.\ assumption violation caused by DA.

A second, parallel line of papers~\cite{Wilson2020BayesianDL,Izmailov2021WhatAB,Kapoor2022OnUT}, focuses more on the link between tempering and aleatoric uncertainty in the data. They claim likelihood tempering can be viewed as adjusting for general model mismatch, and thus \textit{``[...] it would be highly surprising if [$\lambda_{L}=1$] were in fact the best setting,''} and \textit{``we would advocate tempering for essentially any model, [...]. Every model is misspecified}~\cite{Wilson2020BayesianDL}\textit{.''} A view well supported by our own conclusions in Sec.~\ref{sec:analysis}. Next, they show that the cold-posterior effect vanishes in all of the examples of~\cite{Wenzel2020HowGI} if DA is removed~\cite{Izmailov2021WhatAB}. They note that posterior tempering can still be useful by showing that decreasing temperature can improve robustness. In~\cite{Kapoor2022OnUT}, after concluding that misspecified priors are not responsible for the cold-posterior effect, purposeful prior modification~\cite{Fortuin2021BayesianNN} is mentioned as a possible solution. 
Likelihood tempering is interpreted as a way to express how much label noise is believed to be in the data. Their later work~\cite{Marek_Paige_Izmailov_2024}, considering aleatoric uncertainty calibration through the prior, shows the effect to be highly comparable to that of cooling the posterior. A recent paper~\cite{Pitas2023TheFP} ties together the cold-posterior effect literature, providing a comprehensive theoretical and empirical study analyzing their generalization properties via PAC-Bayes bounds. It argues that NLL test-set cross validation is frequentist and will therefore not yield $\lambda_{L}^{*}=1$. In parallel, another paper~\cite{AvellaMedina2022OnTR} shows that for misspecified likelihood, an optimal $\lambda_{L}^{*} < 1$ indeed exists that minimizes the KL divergence between the \textit{``reported''} posterior and the true (but unknown) posterior in Bayesian inference. This provides a rigorous explanation for the empirical success of likelihood-tempered posteriors. 

When we approach the Kalman filtering outlier rejection method in~\cite{Duran-Martin2024OutlierRK} through the lens of posterior tempering, it can be interpreted as adaptively setting $\lambda_{L}<1$ in a time-varying way to down-weigh improbable observations. This makes these results the first to apply tempering ideas to the filtering setting. They find that for the specific scenario where $\lambda_{L}< 1$, recursivity is not broken, and even the closed-form solution is preserved, retaining the computationally efficient Kalman structure. 

\color{black}


\subsection{Our contribution}

Recognizing that the Bayes filter traces its origins to Bayesian inference ideas, and that it is often subject to similar model mismatch, we investigate the merits of tempered posteriors for Bayes filtering.

We propose a canonical posterior-tempering parameterization, comprising both the likelihood- and full-posterior tempering present in the literature. This is then straightforwardly integrated into the Bayes filtering framework, together with a newly-introduced third \emph{belief-tempering} parameter, to yield the parameterized class of \emph{tempered Bayes filters}.

While the likelihood-tempering interpretation of balancing the likelihood with the prior is well-motivated and ubiquitous in the literature, a similarly supported interpretation of full-posterior tempering is still absent. We provide a theoretically grounded interpretation encompassing both modalities. We further provide an analysis of the performance gradient w.r.t.\ the tempered Bayes filter parameters. The results support the view that the \emph{cold-posterior effect}~\cite{Wenzel2020HowGI} is caused by model mismatch in the general sense, agreeing with statements made in~\cite{Wilson2020BayesianDL}. It further enables us to understand under what conditions the method is guaranteed to improve performance. We show that the parameterized class of Bayes filters recovers the maximum a-posteriori (MAP) filter in a specific parameter value limit. The tempering framework can thus alternatively be employed to interpolate between Bayes- and MAP filtering, both special cases of our framework.

We show that tempered Bayes filters admit an efficient online implementation, computationally no more complex than the classic Bayes filter. We further propose an efficient approach to tuning the tempering parameters and demonstrate its merits on a number of numerical studies.

The remainder of this paper is organized as follows. Sec.~\ref{sec:prelims} provides the preliminaries, including notation and a review of Bayes and MAP filtering. Sec.~\ref{sec:main_results} introduces the tempered Bayes filter framework, establishes a set of theoretically informed interpretations, and derives its recursive form. Sec.~\ref{sec:simulations} discusses parameter tuning, presents experimental results on a partially observable grid-world example, and discusses performance. Sec.~\ref{sec:conclusion} concludes the paper with a discussion of implications and future work.

\section{Preliminaries}\label{sec:prelims}

\textit{Notation:}  Let the set of all probability density functions defined over $\mathbb{R}^{n}$ be denoted $\mathcal{P}^{n}:=\{p : \mathbb{R}^{n}\rightarrow\mathbb{R}_{\ge 0} \mid \int_{\mathbb{R}^{n}}p(x)~\mathrm{d}x=1\}$. The set of all probability distributions over a finite alphabet of cardinality $n\in\mathbb{N}$, is further denoted by $\Delta^{n}:=\{p\in\mathbb{R}^{n}_{\ge 0} \mid \sum_{i} p_i=1\}$, i.e., the $(n-1)$-dimensional simplex. Let the $L^{p}$-norm be denoted $\|f\|_{L_p}:=(\int_{X}|f(x)|^{p}\mathrm{d}x)^{1/p}$. All integrals are understood with respect to a base measure $\mu_{X}$ on the state space $X$ and $\mu_{Y}$ on the output space $Y$. For continuous spaces these are the Lebesgue measures, while for finite or countable spaces they are the counting measures. This convention allows all formulas to be written uniformly as integrals, corresponding to finite sums in the discrete case. Only in our discussion on limits, we depart from some of these assumptions and employ the Dirac delta $\delta\left(x-x^*\right)$, to denote all probability mass concentrated at the point $x^{*}$.

\subsection{Hidden Markov Models}

Let the latent state $x_{k}\in X$ at time $k\in\mathbb{N}_{0}:=\mathbb{N}\cup\{0\}$, of a hidden Markov model (HMM) evolve autonomously according to the \emph{transition model} $p(x_{k+1} | x_{k})$, while emitting outputs $y_{k}\in Y$ according to the \emph{output model} $p(y_{k} | x_{k})$, as
\begin{equation}
    p(x_{k+1} | x_{k}):=\operatorname{Prob}(x_{k+1}|x_{k}), \hspace{0.2cm} 
    p(y_{k} | x_{k}):=\operatorname{Prob}(y_{k}|x_{k}). \label{eq:HMM}
\end{equation}
We further know that the initial state $x_0$ of the system is sampled as $x_0\sim p_0$.

\subsection{Filtering}

Given a model $\hat{p}$ approximating the true underlying system behavior $p$, and a sequence of observed outputs $y_{0:k}\in Y^{k+1}$, for $k\in\mathbb{N}_{0}=\mathbb{N}\cup\{0\}$, the goal of a model-based filter is to estimate the true odds of the underlying state $x_{k}$ at time $k$ taking values in $X$ as accurately as possible. The two prevailing filtering approaches in the literature are the \emph{Bayes filter}~\cite{thrun2005probabilistic} and \emph{maximum a posteriori} (MAP) filtering~\cite{Rawlings2009Model}, which we briefly summarize below.

\subsubsection{Bayes Filter}

The Bayes filter iteratively computes the posterior state distribution, given the observations, in two steps. The first step is the application of Bayesian inference to obtain the posterior probability of each trajectory $x_{0:k}\in X^{k+1}$, given the observed data $y_{0:k}\in Y^{k+1}$, often referred to as simply \emph{the posterior}, as
\begin{equation}
p(x_{0:k}|y_{0:k})=\frac{p(y_{0:k}|x_{0:k})p(x_{0:k})}{p(y_{0:k})}, \label{eq:Bayes_posterior}
\end{equation}
where the Markov structure of HMMs~\eqref{eq:HMM} allows us to express the likelihood of the data $y_{0:k}$ given a trajectory $x_{0:k}$, as the product
\[
p(y_{0:k}|x_{0:k})=\prod_{t=0}^{k}p(y_t|x_t),
\]
and the prior distribution of a trajectory $x_{0:k}$, as
\[
p(x_{0:k})=p_{0}(x_{0})\prod_{t=1}^{k}p(x_t|x_{t-1}),
\]
often simply referred to as \emph{the likelihood} and \emph{the prior}, respectively. The probability of observing $y_{0:k}$, often referred to as \emph{the evidence} $p(y_{0:k})$, is \emph{the joint distribution} $p(x_{0:k},y_{0:k})=p(y_{0:k}|x_{0:k})p(x_{0:k})$, marginalized over all trajectories $x_{0:k}\in X^{k+1}$, as $p(y_{0:k})=\int_{X^{k+1}}p(x_{0:k},y_{0:k})\mathrm{d}x_{0:k}$.

The Bayes filter pairs this inference step with a marginalization step to obtain \emph{the posterior over the state} at time $k$ from \emph{the posterior over trajectories}~\eqref{eq:Bayes_posterior}. This yields the \emph{Bayes filter belief} $p(x_{k}|y_{0:k})$, as 
\begin{equation}
p(x_{k}|y_{0:k})=\int_{X^{k}}p(x_{0:k}|y_{0:k})~\mathrm{d}x_{0:k-1},\label{eq:general_Bayes_filter_statement}
\end{equation}
for $x_{k}\in X$.

Integrating over $X^{k}$~\eqref{eq:general_Bayes_filter_statement} is usually prohibitively complex. Instead, by noting that~\eqref{eq:general_Bayes_filter_statement} can be written as
\begin{equation}
\begin{split}
&p(x_{k}|y_{0:k})= \\
&\frac{p(y_{k}|x_{k})\underset{X}{\displaystyle\int}p(x_{k}|x_{k-1})\overbrace{\hspace{-0.2cm}\underset{X^{k-1}}{\displaystyle\int}\hspace{-0.2cm}p(x_{0:k-1}|y_{0:k-1})\mathrm{d}x_{0:k-2}}^{p(x_{k-1}|y_{0:k-1})}\mathrm{d}x_{k-1}}{p(y_{k}|y_{0:k-1})},\label{eq:Bayes_filter_recursion}
\end{split}
\end{equation}
for $x_{k}\in X$, a recursive form is obtained that only requires integration over $X$. Note that $p(y_{k}|y_{0:k-1})$ is an easily computable normalization constant, as it does not depend on $x_{k}\in X$. This recursive  form is the classic Bayes filter.

\subsubsection{Maximum a Posteriori Filter}

While the Bayes filter attempts to approximate the marginal posterior distributions, the \emph{maximum a posteriori (MAP) filter} seeks the \textit{single most probable state sequence} $x_{0:k}^{*}$, conditioned on measurements $y_{0:k}$. The MAP filter thus maximizes, instead of marginalizing, over the posterior~\eqref{eq:Bayes_posterior}, as
\begin{equation}
p(x_{k},x_{0:k-1}^{*}(x_{k})|y_{0:k})=\max_{x_{0:k-1}\in X^{k}}p(x_{0:k}|y_{0:k}),\label{eq:general_MAP_filter_statement}
\end{equation}
for $x_{k}\in X$. The MAP filter admits a similar online recursion to the Bayes filter~\eqref{eq:Bayes_filter_recursion}, as
\begin{equation}
\begin{split}
&p(x_{k},x_{0:k-1}^{*}(x_{k})|y_{0:k})=\\
&\frac{p(y_k | x_{k})\hspace{-0.1cm}\underset{x_{k-1} \in X}{\max} p(x_{k} | x_{k-1}) p(x_{k-1},x_{0:k-2}^{*}(x_{k-1})|y_{0:k-1})}{p(y_{k}|y_{0:k-1})}, \label{eq:MAP_recursion}
\end{split}
\end{equation}
for $x_{k}\in X$. Note that this yields the posterior probability of trajectory $(x_{0:k-1}^{*}(x_{k}),x_{k})$, and not a probability distribution over $x_{k}\in X$.

\subsection{Model Mismatch and Estimator Performance}

Application of the Bayes filter~\eqref{eq:Bayes_filter_recursion} to the true system model $p$ obtains the exact conditional distribution $p(x_{k}|y_{0:k})$. Likewise, the MAP filter~\eqref{eq:MAP_recursion} then obtains the exact most-likely trajectory probability. As the true system model $p$ is generally not available, in practice, a model estimate $\hat{p}$ is substituted for $p$ in the filtering equations~\eqref{eq:Bayes_filter_recursion} and~\eqref{eq:MAP_recursion}. Initializing with $\hat{p}(x_{0}|y_{0})=\hat{p}(y_{0}|x_{0})\hat{p}(x_{0})/\int \hat{p}(y_{0}|x_{0})\hat{p}(x_{0})\mathrm{d}x_{0}$, $k$ iterations of~\eqref{eq:Bayes_filter_recursion} yield the \emph{Bayes filter belief} $b_{\text{B}}(x_{k}|y_{0:k}):=\hat{p}(x_{k}|y_{0:k})$, which approximates the conditional distribution $p(x_{k}|y_{0:k})$, while $k$ iterations of~\eqref{eq:MAP_recursion} yield the \emph{MAP filter belief} $b_{\text{M}}(x_{k}|y_{0:k})\propto\hat{p}(x_{k},x_{0:k-1}^{*}(x_{k})|y_{0:k})$ through normalization over $x_{k}\in X$. For linear dynamic systems with additive Gaussian noise, both these estimators recover the \emph{Kalman filter}.

Accuracy of a belief distribution can be quantified by its expected \emph{negative log-likelihood} (NLL) score~\cite{Wenzel2020HowGI}. The more probability mass an estimator correctly places on the true underlying state value (in expectation), the lower this score. Let the expected NLL score of a filter under consideration, producing normalized belief $b(x_{k}|y_{0:k})=b(x_{k},y_{0:k})/\int_{X}b(x^{\prime}_{k},y_{0:k})\mathrm{d}x^{\prime}_{k}$, for $k\in\mathbb{N}_{0}$, be defined as
\begin{align}
N_{k}:=&\underset{y_{0:k}\in Y^{k+1}}{\underset{x_{0:k}\in X^{k+1}}{\mathbb{E}}}\left[-\log b(x_{k}|y_{0:k})\right],\label{eq:N_k_definition}\\
 = &\hspace{-1mm} \underset{Y^{k+1}}{\int}\hspace{-2mm}p(y_{0:k})\underset{X}{\int}p(x_{k}|y_{0:k})\left[-\log b(x_{k}|y_{0:k})\right]\mathrm{d}x_{k}~\mathrm{d}y_{0:k}, \label{eq:eq:N_k_intermediate}\\
= &\underset{y_{0:k}\in Y^{k+1}}{\mathbb{E}}H_{c}(p(x_{k}|y_{0:k}), b(x_{k}|y_{0:k})),\label{eq:N_k_final}
\end{align}
where $\int_{X^{k}}p(x_{0:k},y_{0:k})\mathrm{d}x_{0:k-1}=p(y_{0:k})p(x_{k}|y_{0:k})$, was used to obtain~\eqref{eq:eq:N_k_intermediate}, and $H_c(p,b):=\int_{X}p(x)[-\log b(x)]\mathrm{d}x$, is the cross-entropy between two distributions. Note that $\log$ represents the natural logarithm here and in the remainder of the work. From the properties of the cross-entropy, it is clear that the NLL score minimum is exactly $\mathbb{E}_{y_{0:k}\in Y^{k+1}}H(p(x_{k} | y_{0:k}))$, achieved at $b=p$, corresponding to the Bayes filter applied to a perfect model, i.e., for $\hat{p}=p$. Whenever $\hat{p}\ne p$, the Bayes filter belief will thus not achieve this minimum. As it turns out, in the context of such imperfect models $\hat{p}$, tempering can improve filter performance as measured by this NLL score. This idea is further developed in the context of Bayes filtering below.

\section{Tempering the Bayes filter} \label{sec:main_results}

Tempered Bayesian inference literature does not generally discern between the two modes of tempering associated with \emph{tempered posteriors}~\cite{Aitchison2020AST,Wilson2020BayesianDL,Nabarro2021DataAI,Noci2021DisentanglingTR,Izmailov2021WhatAB,Bachmann2022HowTF,Kapoor2022OnUT}. While the term \emph{tempered posterior} was originally coined --- in the context of \emph{the cold-posterior effect}, which describes how predictive performance can improve through tempering --- by a paper that tempered the full posterior~\cite{Wenzel2020HowGI}, the majority of tempered posterior papers today temper the likelihood instead.

We will thus use the term \emph{(generally) tempered posterior} to refer to the posterior that results from a general practice of tempering, i.e., by tempering its likelihood, its full-posterior, or both. At the same time, to explicitly discern between the two aforementioned modalities moving forward, we will refer to them as \emph{likelihood tempering} and \emph{full-posterior tempering}, respectively. These two main modalities are easily combined into a fully parameterized canonical posterior tempering framework, which, in the context of~\eqref{eq:Bayes_posterior}, yields the form
\begin{equation}
p_{\lambda}(x_{0:k}|y_{0:k}):\propto\left(p(y_{0:k}|x_{0:k})^{\lambda_{L}}p(x_{0:k})\right)^{\lambda_{P}}, \label{eq:canonically_tempered_posterior}
\end{equation}
where $\lambda_{L}\in\mathbb{R}_{\ge 0}$ is the \emph{likelihood tempering parameter}, $\lambda_{P}\in\mathbb{R}_{\ge 0}$ the 
\emph{posterior tempering parameter}, and equality is obtained after normalization over $x_{0:k}\in X^{k+1}$.

Apart from the two aforementioned posterior-tempering modalities, as inherited from the Bayesian inference tempering literature, the Bayes filter marginalization step~\eqref{eq:general_Bayes_filter_statement} introduces the opportunity for another unique \emph{belief tempering} modality. Substitution of the tempered posterior~\eqref{eq:canonically_tempered_posterior} into the classic Bayes filter belief equation~\eqref{eq:general_Bayes_filter_statement}, and application of belief tempering as parameterized by $\lambda_{B}\in[0,\infty)$, yields the \emph{tempered Bayes belief} as
\begin{equation}
b_{\lambda}(x_{k}|y_{0:k})\propto\left(\int_{X^{k}}p_{\lambda}(x_{0:k}|y_{0:k})~\mathrm{d}x_{0:k-1}\right)^{\lambda_{B}}.\label{eq:tempered_bayes_belief_definition}
\end{equation}
Belief tempering, as an additional degree of parameterization freedom, provides additional performance gains, especially under small training data conditions as demonstrated in our simulations (Sec.~\ref{sec:simulations}). Let $\lambda:=\begin{bmatrix}\lambda_{L} & \lambda_{P} & \lambda_{B}\end{bmatrix}^{\top}$, such that the tempered Bayes filter belief~\eqref{eq:tempered_bayes_belief_definition} at $\lambda_{0}:=\begin{bmatrix}
    1 & 1 & 1
\end{bmatrix}^{\top}$, recovers the classic Bayes filter. 

In the remainder of this section, we first discuss how changes in each of the tempering parameters in $\lambda$ affect the tempered belief (Sec.~\ref{sec:understanding}). We proceed to show that the tempered belief still admits the classic recursive form, as known from the Bayes filter (Sec.~\ref{sec:recursive}). By deriving an expression for the gradient of the negative log likelihood score w.r.t.\ each of the tempering parameters, we clearly outline when performance can be improved (Sec.~\ref{sec:analysis}). We discuss a straightforward finite state-space implementation, provide an alternative, numerically more stable version, and conclude that their complexity is no larger than that of the original Bayes filter (Sec.~\ref{sec:finite_state}). We finish by discussing the consequences of tempering for Kalman filters (Sec.~\ref{sec:lin_gauss}).

\subsection{Tempering interpretation} \label{sec:understanding}

In this section, we investigate the tempered belief parameterization~\eqref{eq:tempered_bayes_belief_definition}, and its sub-component, the canonically tempered posterior~\eqref{eq:canonically_tempered_posterior}, to explain how varying $\lambda$ affects tempered Bayes filter behavior.

\subsubsection{A complete interpretation of posterior tempering}

While prior works, e.g.,~\cite{Wenzel2020HowGI,Catoni2007PACBayesian,grunwald2017inconsistency}, have already interpreted the effect of likelihood tempering as re-balancing the ELBO objective, we seek to do the same while including full-posterior tempering. We present the following proposition towards a complete interpretation of the posterior-tempering framework.

\begin{proposition}[Regularized ELBO maximization]
The tempered posterior, as defined in~\eqref{eq:canonically_tempered_posterior}, is the unique maximizing argument of the entropy regularized, rebalanced ELBO objective, as
\begin{equation}
\begin{split}
&p_{\lambda}(x_{0:k}|y_{0:k})=\\
&\hspace{0.5cm}\arg\max_{q\in\mathcal{P}} \ \Big\{ \mathbb{E}_{q}\left[\log p(y_{0:k}|x_{0:k})\right]-\frac{1}{\lambda_{L}}\operatorname{KL}(q\|p)\\
&\hspace{5.1cm}+\frac{1}{\lambda_{L}}(\frac{1}{\lambda_{P}}-1)H(q)\Big\},\label{eq:interpretation_full_parameterization}
\end{split}
\end{equation}
where $\operatorname{KL}(q\|p)$ is the Kullback-Leibler divergence of distribution $q(x_{0:k})$ from the prior distribution $p(x_{0:k})$, and $H(q)$ its Shannon entropy. \label{thm:regularized_ELBO}
\end{proposition}

\textit{Proof:} Objective~\eqref{eq:interpretation_full_parameterization} can be written as 
\begin{equation}
\int_{X^{k+1}} q(x_{0:k})L_{q}(x_{0:k},y_{0:k},\lambda)~\mathrm{d}x_{0:k},
\label{eq:first_thm1_proof_equation}
\end{equation}
where
\[
\begin{split}
&L_{q}(x,y,\lambda)\\
&\hspace{8mm}= \log p(y|x)-\frac{1}{\lambda_{L}}\log \frac{q(x)}{p(x)}-\frac{1}{\lambda_{L}}(\frac{1}{\lambda_{P}}-1)\log q(x), \\
&\hspace{21.6mm}= -\frac{1}{\lambda_{L}\lambda_{P}}\log\frac{q(x)/C(y,\lambda)}{p(y|x)^{\lambda_{L}\lambda_{P}}p(x)^{\lambda_{P}}/C(y,\lambda)}.
\end{split}
\]
Setting $C(y,\lambda)=\int(p(y|x)^{\lambda_{L}}p(x))^{\lambda_{P}}\mathrm{d}x$, and using definition~\eqref{eq:canonically_tempered_posterior} then yields
\[
L_{q}(x,y,\lambda) = -\frac{1}{\lambda_{L}\lambda_{P}}\left(\log\frac{q(x)}{p_{\lambda}(x|y)}-\log C(y,\lambda)\right).
\]
Substitution back into~\eqref{eq:first_thm1_proof_equation} and using the constraint that $q$ must be a valid probability distribution then allows us to write the objective of~\eqref{eq:interpretation_full_parameterization} as
\begin{equation}
-\frac{1}{\lambda_{L}\lambda_{P}}\operatorname{KL}\left(q\|p_{\lambda}(\cdot|y_{0:k})\right) +\underbrace{\frac{1}{\lambda_{L}\lambda_{P}}\log C(y_{0:k},\lambda)}_{\text{constant w.r.t. }q},
\label{eq:objective_rewritten_as_KL}
\end{equation}
From the properties of the KL-divergence~\cite{Cover1991ElementsOI}, we clearly have that $q^{*}(x)=p_{\lambda}(x|y)$ is the unique maximizer of objective~\eqref{eq:objective_rewritten_as_KL}.\hfill $\square$

Inspecting~\eqref{eq:interpretation_full_parameterization} for the pure likelihood-tempering scenario ($\lambda_{P}=1$), recovers the well-known rebalanced ELBO maximization. As often mentioned in Bayesian inference likelihood-tempering literature, this should be understood as trading off the influence of the data model with that of the prior, i.e., regulating the magnitude of the effect of the data on the posterior. On top of that, Prop.~\ref{thm:regularized_ELBO} enables an interpretation of pure full-posterior tempering ($\lambda_{L}=1$), as entropy regularization. The maximizing argument of~\eqref{eq:interpretation_full_parameterization}, i.e., the tempered posterior~\eqref{eq:canonically_tempered_posterior}, can be understood as the $\lambda_{P}$-trade off of ``\textit{divergence}'' from the untempered posterior with distribution entropy. Decreasing the full-posterior tempering parameter $\lambda_{P}<1$ thus encourages entropy in the resulting posterior, while increasing it $\lambda_{P}>1$ has the effect of reducing tempered posterior entropy.

\subsubsection{Complete distributional tempering and Gibbs distributions}

Full-posterior tempering and belief tempering can both be understood as forms of \emph{complete} (rather than partial) \emph{distributional tempering}. The entropy-regularizing interpretation of full-posterior tempering, as described in Prop.~\ref{thm:regularized_ELBO}, can therefore be understood in the context of belief tempering as well. The following lemma provides additional insight in the exact character of this entropy regularization effect, enabling us to analyse, e.g., the behavior in its extremes.

Let the \emph{temperature-scaled Gibbs distribution} be defined as
\begin{equation}
\operatorname{Gibbs}_{T}(L,x) = \frac{e^{-L(x)/T}}{\int_{X} e^{-L(x^{\prime})/T}\mathrm{d}x^{\prime}}, \label{eq:gibbs_definition}
\end{equation}
for some energy function $  L:X\rightarrow\mathbb{R}$, variable $x\in X$, and temperature $T>0$.

\begin{lemma}[Gibbs distribution]
    Complete-distribution tempering by an exponent $0\le \alpha<\infty$ of a distribution $p\in \mathcal{P}$, as
    \begin{equation}
    \rho_{\alpha}(x) = \frac{p(x)^{\alpha}}{\int p(x^{\prime})^{\alpha}~\mathrm{d}x^{\prime}},\label{eq:complete-distribution tempering}
    \end{equation}
    for $i\in\{1,2,\dots,n\}$, yields the temperature-scaled Gibbs distribution of the associated energy $L(x)=-\log p(x)$, with temperature $T=1/\alpha$, i.e.,
    \[
    \rho_{\alpha}(x)=\operatorname{Gibbs}_{1/\alpha}(-\log p,x).
    \]
    \label{lem:completedistributiontempering_as_gibbs}
\end{lemma}

\textit{Proof:} As $p$ only takes non-negative values, the exponent of its log recovers its original value, allowing us to rewrite~\eqref{eq:complete-distribution tempering} as
\[
\rho_{\alpha}(x)=\frac{e^{\alpha\log p(x)}}{\int e^{\alpha\log p(x^{\prime})}~\mathrm{d}x^{\prime}},
\]
where substitution of $T=1/\alpha$, and $L(x)=-\log p(x)$ yields the exact temperature-scaled Gibbs distribution form~\eqref{eq:gibbs_definition}.\hfill $\square$

The low-temperature limit $T\downarrow 0$ (i.e., $\alpha\to \infty$) of the Gibbs distribution is known to concentrate all probability mass on the minimizer $x^{*}$ (assuming it is unique) of the energy $L(x)$~\cite{Georgi1979CanonicalGM}. Combining this with monotonicity of the log-function leads to
\[
\lim_{\alpha\to\infty}\rho_{\alpha}(x)=\delta(x-x^{*}).
\]
We can thus use Lemma~\ref{lem:completedistributiontempering_as_gibbs} in the context of full-posterior- and belief tempering to conclude
\[
\lim_{\lambda_{P}\to\infty}p_{\lambda}(x_{0:k}|y_{0:k}) = \delta(x_{0:k}-x_{0:k}^{*}),
\]
where $x_{0:k}^{*}=\arg\max_{x_{0:k}\in X^{k+1}} p(y_{0:k}|x_{0:k})^{\lambda_{L}}p(x_{0:k})$, and
\[
\lim_{\lambda_{B}\to\infty}b_{\lambda}(x_{k}|y_{0:k}) = \delta(x_{k}-x_{k}^{*}),
\]
where $x_{k}^{*}=\arg\max_{x_{k}\in X} \int p_{\lambda}(x_{0:k}|y_{0:k})\mathrm{d}x_{0:k-1}$.

\subsubsection{Recovering the MAP filter}

Here, we consider the special case $\lambda_{B}=1/\lambda_{P}$, interpret it as an $L_{p}$-norm and show that in its limit, we recover the MAP-filter based belief $b_{\text{M}}(x_{k}|y_{0:k})$.

\begin{lemma}[$\lambda_{B}=1/\lambda_{P}$ yields $L_{p}$-norm]
    For $\bar{p}=\lambda_{P}=1/\lambda_{B}$, the tempered Bayes filter belief $b_{\lambda}(x_{k}|y_{0:k})$, for $x_{k}\in X$, $y_{0:k}\in Y^{k+1}$, is the normalized $L_{\bar{p}}$-norm over $x_{0:k-1}\in X^{k}$ of the likelihood-tempered posterior distribution $p_{\lambda}(x_{0:k}|y_{0:k})|_{\lambda_{P}=1}$, as
    \[
    b_{\lambda}(x_{k}|y_{0:k})\mid_{\lambda_{B}=\frac{1}{\bar{p}},\lambda_{P}=\bar{p}}\propto\|p_{\lambda}(x_{0:k-1},x_{k},y_{0:k})\mid_{\lambda_{P}=1}\|_{L_{\bar{p}}}.
    \]\label{lem:Lp_norm}
\end{lemma}

\textit{Proof:} This can be seen by the simple substitution of $\bar{p}=\lambda_{P}=1/\lambda_{B}$, and~\eqref{eq:canonically_tempered_posterior} into~\eqref{eq:tempered_bayes_belief_definition}, to yield
\[
b_{\lambda}(x_{k}|y_{0:k})\propto\left(\underset{X^{k}}{\int}(p_{\lambda}(y_{0:k}|x_{0:k})^{\lambda_{L}}p(x_{0:k}))^{\bar{p}}~\mathrm{d}x_{0:k-1}\right)^{1/\bar{p}}\hspace{-2mm},
\]
for $x_{k}\in X$.\hfill $\square$

\begin{theorem}[Recovering the MAP-filter based belief]
In the limit $\lambda_{P}=1/\lambda_{B}\to \infty$, at $\lambda_{L}=1$, the tempered Bayes filter recovers the MAP-filter belief, as
    \[
    \lim_{\lambda_{P}=1/\lambda_{B}\to\infty}b_{\lambda}(x_{k}|y_{0:k})\mid_{\lambda_{L}=1} = b_{\text{M}}(x_{k}|y_{0:k}).
    \]\label{thm:MAP}
\end{theorem}

\textit{Proof:} Using Lemma~\ref{lem:Lp_norm}, we can take the definition of the tempered Bayes belief~\eqref{eq:tempered_bayes_belief_definition}, and substitute $\bar{p}$ for $\lambda_{P}$, $1/\bar{p}$ for $\lambda_{B}$, and $1$ for $\lambda_{L}$ to obtain
\begin{equation}
\begin{split}
&b_{\lambda}(x_{k}|y_{0:k})\mid_{\lambda_{B}=\frac{1}{\bar{p}},\lambda_{P}=\bar{p},\lambda_{L}=1}\\
&\hspace{2cm}\propto\|p_{\lambda}(x_{0:k-1},x_{k},y_{0:k})\mid_{\lambda_{P}=\lambda_{L}=1}\|_{L_{\bar{p}}}\\
&\hspace{2cm}=\|p(x_{0:k-1},x_{k},y_{0:k})\|_{L_{\bar{p}}},
\end{split}\label{eq:Lp_proof_setup}
\end{equation}
where we can then take the limit $\bar{p}\to\infty$ on both sides. We know from the properties of the $L_{p}$-norm that the right-hand side becomes
\begin{equation}
\begin{split}
&\lim_{\bar{p}\to\infty}\|p(x_{0:k-1},x_{k},y_{0:k})\|_{L_{\bar{p}}}= \|p(x_{0:k-1},x_{k},y_{0:k})\|_{L_{\infty}} \\
&\hspace{4cm}= \max_{x_{0:k-1}\in X}|p(x_{0:k-1},x_{k},y_{0:k})|,
\end{split} \label{eq:Linfinity_limit}
\end{equation}
where the absolute marks may be ignored as the argument is non-negative. Substitution of~\eqref{eq:Linfinity_limit} into~\eqref{eq:Lp_proof_setup} then yields the exact form~\eqref{eq:general_MAP_filter_statement}.\hfill $\square$

Next, we show that the tempered Bayes filter admits a recursive form for all $\lambda\in[0,\infty)^{3}$, allowing us to avoid computing the marginalization integral~\eqref{eq:general_Bayes_filter_statement}.

\subsection{Recursive form}\label{sec:recursive}

It is well known that the Bayes filter and the MAP filter admit a recursive implementation. Below, we show that the tempered Bayes filter too admits a recursive form for all $\lambda\in[0,\infty)^{3}$.

\begin{proposition}[Tempered Bayes filter recursive form]
    The tempered Bayes filter belief at time $k\in\mathbb{N}$ can be computed by an integral over only the previous state when the tempered Bayes filter belief at time $k-1$ is known, as
    \begin{equation}
    \begin{split}
    &b_{\lambda}(x_{k}|y_{0:k})\propto\Big(p(y_{k}|x_{k})^{\lambda_{L}\lambda_{P}} \\
    &\hspace{1cm}\cdot\underset{X}{\int}p(x_{k}|x_{k-1})^{\lambda_{P}}b_{\lambda}(x_{k-1}|y_{0:k-1})^{1/\lambda_{B}}~\mathrm{d}x_{k-1}\Big)^{\lambda_{B}}\hspace{-2mm}.
    \end{split}\label{eq:theorem_recursive_belief_equation}
    \end{equation}
\end{proposition}

\textit{Proof:} We recall the tempered Bayes filter belief definition, as
\begin{equation}
b_{\lambda}(x_{k}|y_{0:k})\propto\left(\int_{X^{k}}\left(p(y_{0:k}|x_{0:k})^{\lambda_{L}}p(x_{0:k})\right)^{\lambda_{P}}\mathrm{d}x_{0:k-1}\hspace{-1mm}\right)^{\lambda_{B}}\hspace{-2mm},\label{eq:belief_for_proof}
\end{equation}
having exact equality when the right-hand side is normalized over $x_{k}\in X$. We may rewrite the term inside the integral as
\[
\begin{split}
\left(p(y_{0:k}|x_{0:k})^{\lambda_{L}}p(x_{0:k})\right)^{\lambda_{P}}\hspace{4.6cm}&\\
=\left((p(y_{k}|x_{k})p(y_{0:k-1}|x_{0:k-1}))^{\lambda_{L}}p(x_{k}|x_{k-1})p(x_{0:k-1})\right)^{\lambda_{P}}&\\
\propto p(y_{k}|x_{k})^{\lambda_{L}\lambda_{P}}p(x_{k}|x_{k-1})^{\lambda_{P}}p_{\lambda}(x_{0:k-1}|y_{0:k-1}),&
\end{split}
\]
which we substitute back into~\eqref{eq:belief_for_proof}, to yield exactly~\eqref{eq:theorem_recursive_belief_equation}, where the independence of $p(y_{t}|x_{t})$ to $x_{0:k-1}$, and $p(x_{k}|x_{k-1})$ to $x_{0:k-2}$, have been used to take them out of the $x_{0:k-1}$, and $x_{0:k-2}$ integrals respectively.\hfill $\square$

\subsection{Gradient Analysis} \label{sec:analysis}

We analyze the effect of $\lambda$ on the tempered Bayes filter performance as quantified by the NLL score~\eqref{eq:N_k_definition}. to show that under mismatch $\hat{p}\ne p$, and a few mild assumptions, the derivative $d N_{k}/d\lambda |_{\lambda_{0}}\ne 0$, i.e., tuning $\lambda\ne \lambda_{0}$ improves upon the $\lambda=\lambda_{0}$ (classic Bayes filter) score. 

\begin{theorem}[NLL Derivative w.r.t.\ $\lambda$]
The gradient of the negative log-likelihood score~\eqref{eq:N_k_definition} of the tempered Bayes filter belief at time $k\in\mathbb{N}$, w.r.t.\ $\lambda$, is
\begin{equation}
\begin{split}
&\nabla_{\lambda} N_{k}(\lambda)=\\
&\hspace{4mm}\underset{y_{0:k}\in Y^{k+1}}{\mathbb{E}}\int_{X}(b_{\lambda}(x_{k}|y_{0:k})-p(x_{k}|y_{0:k}))\ell_{\lambda}(x_{k},y_{0:k})~\mathrm{d}x_{k},
\end{split} \label{eq:NLL_der_general}
\end{equation}\label{thm:derivative}
\end{theorem}
where 
\[
\ell_{\lambda}(x_{k},y_{0:k}) = \underset{\hat{p}_{\lambda}(x_{0:k-1}|x_{k},y_{0:k})}{\mathbb{E}}\log \begin{bmatrix}
\hat{p}(y_{0:k}|x_{0:k})^{\lambda_{P}\lambda_{B}} \\
\hat{p}_{\lambda}(x_{0:k},y_{0:k})^{\lambda_{B}/\lambda_{P}} \\
b_{\lambda}(x_{k},y_{0:k})^{1/\lambda_{B}}
\end{bmatrix}.
\]

\textit{Proof:} See Appendix.\hfill $\square$

Evaluating this derivative~\eqref{eq:NLL_der_general} at $\lambda=\lambda_{0}$, yields
\[
\begin{split}
&\nabla_{\lambda} N_{k}(\lambda)\Big|_{\lambda=\lambda_{0}}=\\
&\hspace{3mm}\underset{y_{0:k}\in Y^{k+1}}{\mathbb{E}}\int_{X}(\hat{p}(x_{k}|y_{0:k})-p(x_{k}|y_{0:k}))\ell_{\lambda_{0}}(x_{k},y_{0:k})~\mathrm{d}x_{k}.
\end{split}
\]
The derivative at $\lambda=\lambda_{0}$ is thus clearly identically zero when the model $\hat{p}$ is perfect ($\hat{p}=p$); agreeing directly with the theoretical understanding that the classic Bayes filter is the optimal estimator when having access to a perfect model. 

We can re-interpret~\eqref{eq:NLL_der_general} as the difference in the expected value of model properties $\ell_{\lambda_{0}}(x_{k},y_{0:k})$ w.r.t.\ the model $\hat{p}$, and the true system $p$. Getting this to zero requires tuning of $\lambda$ until these expected values agree. While we do not analytically show under what conditions such a static point exists, in our simulation results, under natural model mismatch conditions, such point is always found.

For deterministic models $\hat{p}$, either states and outputs with zero probability (according to the model) are never encountered and the derivative is zero (the model would thus need to be perfect), or a zero-probability element inside the log term does occur with non-zero probability according to the expected value, resulting in an ill defined derivative. Deterministic models, whether perfectly identified or not, can thus not benefit from tempering. In contrast, imperfect models identified with a uniform (or at least having full support) prior, in the vast majority of cases, can benefit from tuning $\lambda\ne \lambda_{0}$ (outside a set of highly specific conditions), as for $p\ne \hat{p}$ and $\hat{p}$ not deterministic, the first derivative is generally not zero. We further note that the sign of the derivative depends on the \emph{entropy of the paths as described by the identified model} $\hat{p}$, i.e., the backward uncertainty (under $\hat{p}$) about which past trajectory $x_{0:k-1}$ could have led to the current state $x_{k}$, given observations $y_{0:k}$.

\subsection{Finite-space systems} \label{sec:finite_state}

For finite state- and output space systems, no strong assumptions are needed for~\eqref{eq:theorem_recursive_belief_equation} to be computable. This allows us to provide a general purpose algorithm for this setting. To prevent repetition of identical calculations, we propose employing a pre-computed \emph{tempered model}, comprising of
\[
\hat{p}_{\lambda}(x^{\prime}|x):=\hat{p}(x^{\prime}|x)^{\lambda_{P}}, \ \ \ \hat{p}_{\lambda}(y|x):=\hat{p}(y|x)^{\lambda_{L}\lambda_{P}\lambda_{B}},
\]
and $\hat{p}_{0,\lambda}(x_0):=\hat{p}_0(x_0)^{\lambda_{P}}$, for $x,x^{\prime},x_{0}\in X$, $y\in Y$, and $X$ and $Y$ finite.

By initializing $b_0(x_0)\propto\hat{p}_{\lambda}(y_{0}|x_0)^{\lambda_{B}}\hat{p}_{0,\lambda}(x_0)^{\lambda_{B}}$, for all $x_{0}\in X$, and then recursively computing
\[
\begin{split}
&b_{k}(x_{k})\propto\\
&\hspace{0.8cm}\hat{p}_{\lambda}(y_{k}|x_{k})\hspace{-1mm}\left(\sum_{x_{k-1}\in X}\hat{p}_{\lambda}(x_{k}|x_{k-1})b_{k-1}(x_{k-1})^{1/\lambda_{B}}\right)^{\lambda_{B}}\hspace{-2mm},
\end{split}
\]
for all $x_{k}\in X$, where generally the proportionality statements would be resolved to equality by a simple normalization at each time step, we obtain the tempered Bayes belief at time $k\in\mathbb{N}$. This simple approach has a computational complexity of $O(k|X|^{2})$ --- or a mere $O(|X|^{2})$ when taking the recursive perspective --- which can be reduced further by the use of sparse algebra when many elements of the tempered transition model $\hat{p}_{\lambda}(x^{\prime}|x)$ are zero.

A straightforward implementation of the above algorithm can be very sensitive to numerical underflow in applications with many low probabilities, and high values of $\lambda$. For applications sensitive to such issues, we provide a numerically more stable alternative Alg.~\ref{alg:temperedBayesImproved}. The introduction of stability measures does not increase the (asymptotic) computational complexity, but does yield a moderate constant-factor slowdown which depends on the hardware and sparsity.

\begin{algorithm}[h!]
\SetKwFunction{Flogsumexp}{logsumexp}
\SetKwProg{Fn}{Function}{}{end}
\hrule\vspace{2pt} 
\hrule\vspace{2pt} 
\caption{Tempered Bayes filter in finite spaces --- Numerically stable implementation}\label{alg:temperedBayesImproved}
\textbf{Input} $\hat{p}_{0}$, $\hat{p}(x^{\prime}|x)$, $\hat{p}(y|x)$, $\lambda_{L}$, $\lambda_{P}$, $\lambda_{B}$, $k$ 

\vspace{1mm}
\Fn{\Flogsumexp{$v$}}
        {
        $m = \max(v)$
        
        $\nu = m + \log \sum \exp (v-m)$

        \Return $\nu$
        }

\BlankLine
\tcc{Initialization step}

$P_0(x) =\lambda_{P}\log \hat{p}_{0}(x), \ \ \forall x\in X$

$A_{x^{\prime}}(x) =\lambda_{P}\log \hat{p}(x^{\prime}|x), \ \ \forall x\in X, x^{\prime}\in X$

$C_{y}(x) =\lambda_{L}\lambda_{P}\log \hat{p}(y|x), \ \ \forall x\in X, y\in Y$

\BlankLine

\tcc{Algorithm}

$\ell_0 = C_{y_0}+P_0$ 

$\bar{\ell}_0 = \ell_0 - \Flogsumexp(\ell_0)$ 

\For{ $t \in \{1,2,\dots,k\}$ }
    {
    \For{ $x^{\prime}\in X$ }
        {
        $e_{t}(x^{\prime}) = \Flogsumexp(A_{x^{\prime}}+\bar{\ell}_{t-1})$
        }
        
    $\ell_{t}=C_{y_t}+e_{t}$
    
    $\bar{\ell}_{t}=\ell_{t} - \Flogsumexp(\ell_{t})$
    }
    
    \Return$\bar{b}_{k}=\operatorname{exp}(\lambda_{B}\bar{\ell}_{k}-\Flogsumexp(\lambda_{B}\bar{\ell}_{k}))$.
    
    \vspace{1mm}
\hrule\vspace{2pt} 
\hrule\vspace{2pt} 
\end{algorithm}

\subsection{Tempered Kalman filter} \label{sec:lin_gauss}

Let us consider the class of linear-Gaussian HMMs to explain how tempering affects the Kalman filter. For linear systems subject to additive Gaussian disturbances and measurement noise
\[
x_{k+1}=Ax_{k}+w_{k}, \ \ \ y_{k} = Cx_{k} + v_{k},
\]
where $w_{k}\sim\mathcal{N}(0,\Sigma_{w})$, $v_{k}\sim\mathcal{N}(0,\Sigma_{v})$, and $x_0\sim \mathcal{N}(\overline{x}_0,\Sigma_{x_0})$, we obtain a Gaussian prediction model $p(x_{k}|x_{k-1})= \mathcal{N}(x_{k};Ax_{k-1},\Sigma_{w})$, a Gaussian output model $p(y_{k}|x_{k})= \mathcal{N}(y_{k};Cx_{k},\Sigma_{v})$, and a Gaussian initial state distribution $p_0(x_0) = \mathcal{N}(x_{0};\overline{x}_{0},\Sigma_{x_0})$.

Through an induction argument, we show that keeping track of only the (tempered) mean and (tempered) covariance is sufficient to run a tempered Kalman filter. The tempered Bayes belief is initialized as
\[
\begin{split}
b_{\lambda}(x_{0}|y_{0})&\propto (p(y_0|x_0)^{\lambda_{L}}p_0(x_0))^{\lambda_{P}\lambda_{B}}\\
&\propto\mathcal{N}(x_0;\mu_{\lambda,0|0},\Sigma_{\lambda,0|0}),
\end{split}
\]
where, using Gaussian multiplication rules, we find
\[
\Sigma_{\lambda,0| 0}=\frac{1}{\lambda_{P}\lambda_{B}}\left(\Sigma_{x_0}^{-1}+\lambda_L C^{\top} \Sigma_v^{-1} C\right)^{-1},
\]
and
\[
\mu_{\lambda,0| 0}=\Sigma_{0 | 0}\left(\lambda_P \lambda_B \Sigma_{x_0}^{-1} \bar{x}_0+\lambda_L \lambda_P \lambda_B C^{\top} \Sigma_v^{-1} y_0\right),
\]
where $\lambda=\lambda_{0}$ can be shown to recover the standard Kalman filter equations, and $\lambda_{L}$ is clearly governing the predictions (prior) versus observations (likelihood) trade-off.

Assuming $b_{\lambda}(x_{k-1}|y_{0:k-1})=\mathcal{N}(x_{k-1};\mu_{\lambda,k-1|k-1},\Sigma_{\lambda,k-1|k-1})$, we find $b_{\lambda}(x_{k}|y_{0:k})$, using~\eqref{eq:theorem_recursive_belief_equation}, as
\[
b_{\lambda}(x_{k}|y_{0:k})=\mathcal{N}\left(x_k ; \mu_{\lambda, k | k}, \Sigma_{\lambda, k | k}\right),
\]
where
\[
\Sigma_{\lambda, k | k}=\frac{1}{\lambda_{P}\lambda_{B}}((\lambda_{P}\lambda_{B}\Sigma_{\lambda, k | k-1})^{-1}+\lambda_L C^{\top} \Sigma_v^{-1} C)^{-1}\hspace{-4mm},
\]
where
\[
\Sigma_{\lambda, k | k-1}=A \Sigma_{\lambda, k-1 | k-1} A^{\top}+\frac{1}{\lambda_{P}\lambda_{B}} \Sigma_w,
\]
and
\[
\begin{split}
&\mu_{\lambda, k | k}=\\
&\hspace{0.2cm}\Sigma_{\lambda, k | k}\left(\Sigma_{\lambda, k | k-1}^{-1} A \mu_{\lambda, k-1 | k-1}+\lambda_L \lambda_P \lambda_B C^{\top} \Sigma_v^{-1} y_k\right),
\end{split}
\]
we conclude that if $b_{\lambda}(x_{k-1}|y_{0:k-1})$ is Gaussian, $b_{\lambda}(x_{k}|y_{0:k})$ is also Gaussian. Combined with the finding that for $k=0$, the tempered Bayes belief is Gaussian, we conclude that for all $k\in\mathbb{N}$, the tempered Bayes belief will be Gaussian.

We further conclude that on the line $\lambda_{P}=1/\lambda_{B}$, the actual value of $\lambda_{P}$ (and the associated $\lambda_{B}=1/\lambda_{P}$) does not affect the mean for any $k\in \mathbb{N}$. This corresponds well with the finding that the Bayes filter and the MAP filter, for $\lambda_{L}=1$, are on this line. The fact that moving over the line by varying $\lambda_{P}$ does not affect the mean, also corresponds well with the known fact that the Kalman filter state estimate is equivalently obtained by both the Bayes and MAP filter approaches.

\section{Simulations}\label{sec:simulations}

To investigate the behavior of our tempered Bayes filter in practice, we have constructed a stochastic partially observable grid-world example. An autonomous agent, e.g., a mobile robot, quadcopter, or marine drone, initializes with 50-50 odds in either (i) a stochastically perturbed (as in windy, slippery, or crowded) region of the state space, or (ii) a deterministic (calm) region. From there, depending on in which region it initialized, it either (i) stochastically or (ii) deterministically moves towards the home state, where it stays indefinitely once arrived. We do not know where it has initialized, but we can use sensor measurements, e.g., radar, lidar, or gps signals, subject to significant measurement noise, to estimate its location. 

The system model with $X=\{1,2,\dots,n_{x}\}$, $n_{x}=39$, and $x_{\text{home}}=20$, thus has $p(x_0) = 0.5$ for $x_{0} \in\{1,n_{x}\}$, zero otherwise. Let $X_{\text{stoch}}=\{1,2,\dots,x_{\text{home}}-1\}$, and $X_{\text{det}}=\{x_{\text{home}}+1,x_{\text{home}}+2,\dots,n_{x}\}$. Its dynamics, for $k\in\{0,1,\dots,h\}$, are then described by
\begin{equation}
p(x_{k+1}|x_{k}) = \begin{cases}
    1 & \text{if } x_{k}=x_{\text{home}}, \text{ and } x_{k+1} = x_{\text{home}}, \\
    1 & \text{if } x_{k}\in X_{\text{det}}, \hspace{2mm} \text{ and }x_{k+1}=x_{k}-1,\\
    0.1 & \text{if } x_{k}\in X_{\text{stoch}}, \text{ and }x_{k+1}=x_{k}+3,\\
    0.15 & \text{if } x_{k}\in X_{\text{stoch}}, \text{ and }x_{k+1}=x_{k}+2,\\
    0.5 & \text{if } x_{k}\in X_{\text{stoch}}, \text{ and }x_{k+1}=x_{k}+1,\\
    0.15 & \text{if } x_{k}\in X_{\text{stoch}}, \text{ and }x_{k+1}=x_{k},\\
    0.1 & \text{if } x_{k}\in X_{\text{stoch}}, \text{ and }x_{k+1}=x_{k}-1,\\
    0 & \text{otherwise}.
\end{cases}\label{eq:stoch_region}
\end{equation}
The probabilities in the stochastic region~\eqref{eq:stoch_region} are further modified to ensure all probability mass placed on $x_{k+1}<1$ and $x_{k+1}>x_{\text{home}}$ is instead added to $x_{k+1}=1$ and $x_{k+1}=x_{\text{home}}$, respectively. The observations of the system are subject to zero-mean Gaussian noise with a variance as $p(y_k|x_k)=\mathcal{N}(x_k,(n_{x}/8)^{2})$.

\subsection{Tuning the parameters}\label{sec:tuning}

The tempered Bayes filter most straightforwardly enables filtering performance improvement in scenarios where only a limited amount of \emph{``labeled''} data is available of state- and output trajectories. We propose the following training procedure. The training set is split into $K>1$ folds. For each $i\in\{1,2,\dots,K\}$, we thus have a \emph{training subset} (consisting of $K-1$ training set folds) and an associated \emph{validation subset} (the remaining fold from the training set). We identify a model $\hat{p}_i$ on training subset $i$ (e.g., a uniform-prior maximum-likelihood model estimate) using the same technique that will be used to identify our final model $\hat{p}$ on the entire training set. Each training subset model $\hat{p}_i$ is then used to apply and score a tempered Bayes filter on the associated validation subset data. This score is subsequently optimized by gradient descent of $\lambda$. This yields $K$ optimal generalizing parameterizations $\lambda$, which we average to obtain $\lambda^{*}$. The tempered Bayes filter that results from this procedure thus consists of $\hat{p}$ as identified on the entire training set, applied in a tempered Bayes filter with $\lambda=\lambda^{*}$. Note that, due to the NLL score becoming infinite whenever a test-set observation or state is encountered that the filter gives zero probability mass, it is advised to use a model $\hat{p}$ identification method that gives (small) non-zero probability even to outcomes not observed in the data, through, e.g., a uniform prior.

\subsection{Experimental results}

Instead of having access to the parameters describing the system model, we assume only access to $N\in\mathbb{N}$ state- and output trajectories. The data-set $D=\{(x_{0:h},y_{0:h})_{i=1}^{N}\}\in (X\times Y)^{hN}$, is thus all that is available to construct our filter. The tuning mechanism from Sec.~\ref{sec:tuning} is then applied on this data-set with a $70:30$ training-test split and $K=5$ folds on the training set. 

\begin{figure}[tb]
    \centering
    \includegraphics[width=8cm]{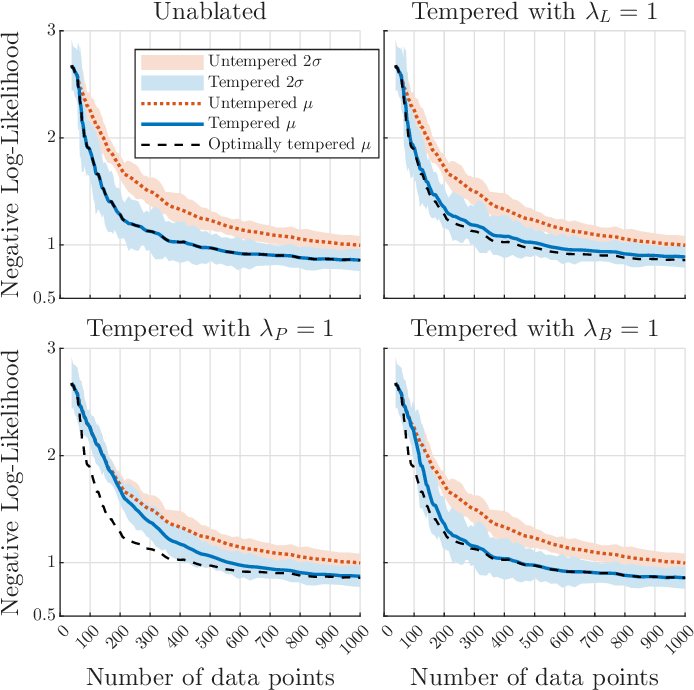}
    \caption{Filtering performance as a function of the amount of available data used for identification under unablated (standard) and ablated conditions.}
    \label{fig:ablation_NLL}
\end{figure}
    
\begin{figure}[tb]
    \centering
    \includegraphics[width=8cm]{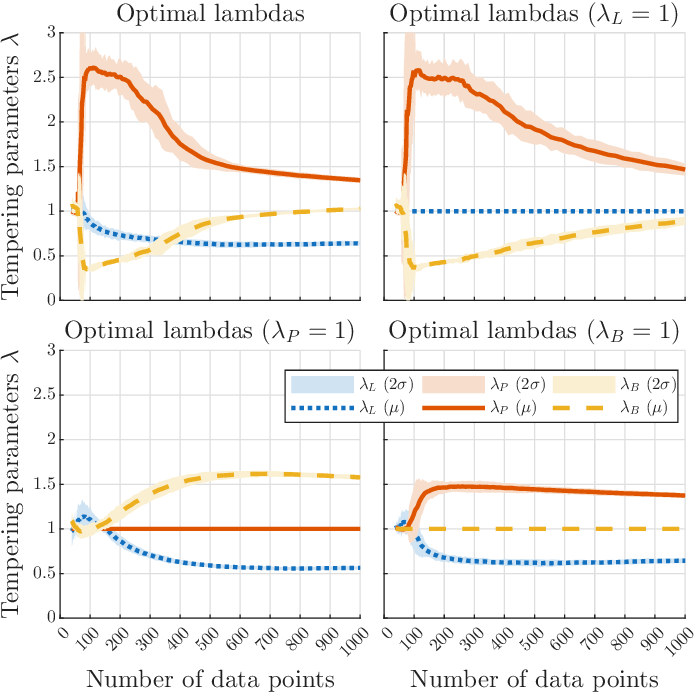}
    \caption{Tuned lambda values as a function of the amount of available data under unablated (standard) and ablated conditions.}
    \label{fig:ablation_lambda}
\end{figure}

The resulting behavior has been investigated as a function of the data-set size. One hundred unique data-set sizes between $N=n_{x}=39$, and $N=1000$ have been simulated for twenty different random seeds each. The results are displayed in Figures~\ref{fig:ablation_NLL}~and~\ref{fig:ablation_lambda}, alongside the (untempered) classic Bayes filter performance in orange.

\subsubsection{Tempered Bayes filter performance}

In the unablated case (standard tempered Bayes filter), found in the second quadrant of Fig.~\ref{fig:ablation_NLL}, it becomes clear that estimation performance is most benefited by tempering for small- to medium sized data-sets. When the size of the data-set is too limited, i.e., $N\approx n_{x}$, the model is so weak that tempering is not able to improve performance at all. The moment the size of the data set starts to increase to above this lower-bound, is when the improvement rises the fastest. Conversely, at very large data-set sizes, the tempered method eventually loses its edge as the model becomes less and less imperfect, making the classic Bayes filter more and more optimal.

We further note that for very small data-sets slightly above $N\approx n_{x}$, the tempered Bayes filter converges to a mode akin to the $L_{p}$-norm discussed in Lemma~\ref{lem:Lp_norm}, which, using Thm.~\ref{thm:MAP}, we interpret as interpolating the Bayes and the MAP filter. In the top two figures of Fig.~\ref{fig:ablation_lambda}, the optimal values of $\lambda_{B}$ and $\lambda_{P}$ indeed stay close to $\lambda_{B}=1/\lambda_{P}$. We believe this to be due to the uniform prior used in the identification of $\hat{p}$. For very small data-sets, it is desirable to emphasize the relatively small effect of those few measurements on the model.

\subsubsection{Ablation study}

Figures~\ref{fig:ablation_NLL}~and~\ref{fig:ablation_lambda}, in quadrants I, III, and IV, further describe the effects parameter ablation has on tempered Bayes filter performance. Besides the full tempered Bayes filter, three versions, each with one parameter kept neutral, are displayed to demonstrate the individual impact of each parameter on the resulting performance. We note that omission of $\lambda_{B}$ is most damaging to performance in the smallest data-set size region, while $\lambda_{P}$ affects the middle-sized data-set region most, and the effect of $\lambda_{L}$ instead is most pronounced at slightly larger data-set sizes. The parameter $\lambda_{P}$ turns out to have the largest cumulative beneficial effect on performance. 

By inspecting Fig.~\ref{fig:ablation_lambda}, we further find that $\lambda_{B}$ and $\lambda_{P}$ have a counteracting effect when one is larger- and the other smaller than one (neutral). One may understand this from the fact that constraining $\lambda_{B}=1$ causes $\lambda_{P}$ to take on less extreme values, and that constraining $\lambda_{P}=1$ results in $\lambda_{B}$ rising above neutral to take its place. The product of $\lambda_{B}$ and $\lambda_{P}$ thus floats around the same level above neutral in all figures. This corresponds well with the interpretation from Sec.~\ref{sec:understanding}, stating that full distributional tempering regulates entropy, something that both $\lambda_{P}$ and $\lambda_{B}$ can be understood to effect.

\begin{figure}
    \centering
    \includegraphics[width=8cm]{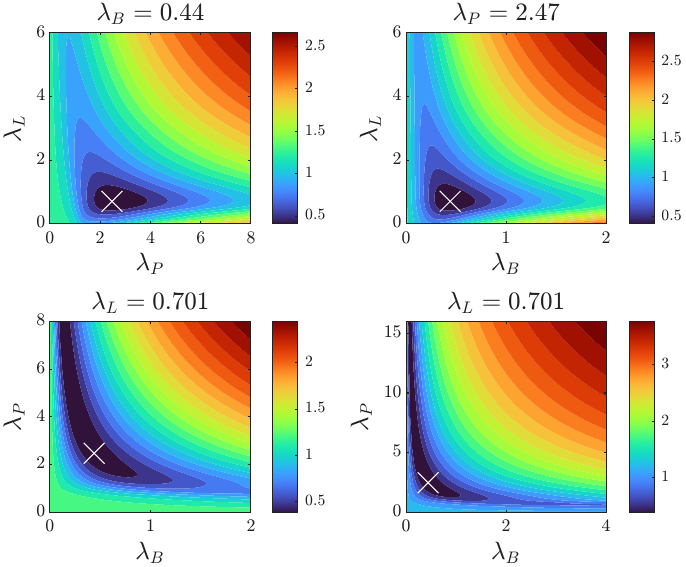}
    \caption{Log of the NLL cost landscape across values of $\lambda$ after training on the 195-data-points data set. The single $\lambda$-parameter value that is kept constant in each of the figures is kept at its associated optimal value.}
    \label{fig:cost_landscape}
\end{figure}

\subsubsection{Cost landscape}

The optimal parameter values have been found through gradient descent. Given the complexity of the cost function~\eqref{eq:N_k_definition}, coupled with the fact that not its expected value, but a Monte-Carlo approximation of its expected value is evaluated, one might wonder whether this is an adequate approach. 

Towards a better understanding of the cost landscape, for $N=195$, the cost landscape has been visualized in Fig.~\ref{fig:cost_landscape}. From these results, we understand that --- for this specific example, but likely more general --- the cost landscape is relatively friendly and well conditioned for gradient-descent optimization. This also strengthens the belief that the lambdas found and displayed in~\ref{fig:ablation_lambda} have a high change of being globally optimal.

\begin{figure}
    \centering
    \includegraphics[width=8cm]{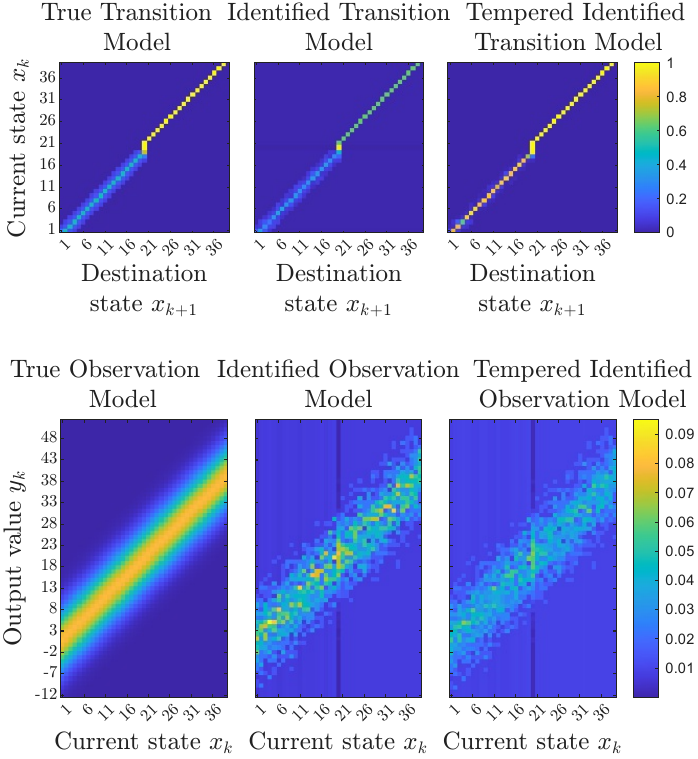}
    \caption{The true, identified and tempered models at the optimal $\lambda^{*}$ at 195 data points.}
    \label{fig:models}
\end{figure}

\subsubsection{The effect of tempering on identified models}

We consider the \emph{tempered} versions of the identified model as defined in Sec.~\ref{sec:finite_state}. Doing so for $N=195$ yields the visualization in Fig.~\ref{fig:models}. 

In the top row, we see how the identified transition model for relatively small data-sets like $N=195$ is far from perfect. After converging to the optimal parameterization, we see that the associated tempered model has emphasized, or \emph{sharpened} the transition probabilities, if you will. 

In the bottom row, the identified observation model seems to deviate even more strongly from the true underlying model than the transition model in the top row does. With more possible outcomes at each time step, less data observations are available per-outcome, making its increased remaining divergence from the true underlying model unsurprising. Surprising is that the parameter tuning yields $\lambda_{L}^{*}<1$. As the regularized ELBO interpretation states that reducing $\lambda_{L}<1$ has the effect of up-regulating the influence of the likelihood model on the resulting posterior.

\section{Conclusion and Discussion}\label{sec:conclusion}

We introduce the \emph{tempered Bayes filter}, a three-parameter extension of the Bayes/MAP filters that unifies \mbox{likelihood-,} \mbox{full-posterior-,} and belief-tempering in a single canonical framework. We provide a theoretically-grounded interpretation through an entropy-regularized ELBO maximization, establish a recursion that preserves the exact Bayes filter computational complexity, and show how it recovers the MAP filter in the limit.

A first-order analysis of filter performance, as measured by the negative log likelihood score, is provided that demonstrates how under general model mismatch, the classic Bayes filter can be improved upon, motivating tempering as a principled calibration mechanism. Algorithmic implementations for the finite-state setting are provided, both a fast- and a numerically more stable version enable online use.

Experiments on a partially observable grid-world example system, identified from limited data, shows consistent improvements over the classic Bayes filter. The distinct role of each tempering parameter and the qualitative shape if their cost landscapes is investigated empirically.

Future work includes non-Gaussian continuous-state extensions, possibly in the direction of particle filters, and an adaptive parameter optimization scheme. 

The proposed framework thus offers a practical, efficient, and theoretically grounded approach to filtering under imperfect models.

\subsection{Consequences for Bayesian inference} \label{sec:soft_inference}

While we have obtained our tempered Bayes filter by applying ideas from Bayesian inference literature to  the Bayes filter setting, a number of conclusions we draw above apply back onto Bayesian inference. Our parameterization~\eqref{eq:canonically_tempered_posterior} is straightforwardly generalized back onto the Bayesian inference setting by considering the \textit{``hidden''} model parameters $\theta$ as the static analogue to our dynamically evolving latent states $x_{0:h}$, and the general concept of data $D_{0:h}$ as analogue to our observations $y_{0:h}$ to yield
\begin{equation}
p(\theta | D) \propto \left(\prod_{k=0}^{h} p(D_{k}|\theta)^{\lambda_{L}}p(\theta)\right)^{\lambda_{P}}. \label{eq:tempered_bayesian_inference}
\end{equation}
Interpretation~\eqref{eq:interpretation_full_parameterization} straightforwardly applies, enabling the entropy-regulating interpretation of Bayesian inference full-posterior tempering $\lambda_{P}$. The log-sum-exp approach from Alg.~\ref{alg:temperedBayesImproved} also requires only little modification to enable numerically stable updates to~\eqref{eq:tempered_bayesian_inference}. Most valuably, the analysis from Thm.~\ref{thm:derivative} applies. Thus enabling the formal understanding that not one or multiple specific methodological errors, but a very general class of model mismatch artifacts result in the benefits of \emph{both} likelihood-, and full-posterior tempering.

\appendix
\section{Appendix}

\subsection{NLL gradient}\label{apx:NLL_gradient}
In this section, we derive analytical expressions for the first derivatives of the NLL score w.r.t.\ the tempering parameters $\lambda_{i}$ for $i\in\{L,P,B\}$. We begin by deriving the components common to all three derivatives, and then address the steps specific to each derivative individually. Afterwards, we combine the three into a single vector-valued expression.

The relevant objects in our derivations, using simplified notation, are
\[
b_{\lambda}(x|y) = \frac{b_{\lambda}(x,y)}{b_{\lambda}(y)}=\frac{b_{\lambda}(x,y)}{\int_{X}b_{\lambda}(x,y)\mathrm{d}x},
\]
and
\[
b_{\lambda}(x,y)=\Bigg(\int_{X^{k}}\hat{p}_{\lambda}(x_{0:k-1},x,y)~\mathrm{d}x_{0:k-1}\Bigg)^{\lambda_{B}},
\]
where
\[
\hat{p}_{\lambda}(x_{0:k-1},x,y) = (\hat{p}(y|x_{0:k-1},x)^{\lambda_{L}}\hat{p}(x_{0:k-1},x))^{\lambda_{P}}.
\]

\subsubsection{Common components}

Let $\lambda_i$, for $i\in\{L,P,B\}$, represent a yet unspecified member of the three tempering parameters. The initial steps towards the NLL derivative do not depend on the choice of parameter and are thus treated in general form here.

Assuming that for each $i\in\{L,P,B\}$, $x\mapsto \log b_{\lambda}(x|y)$ is differentiable in $\lambda_{i}$ for almost every $x\in X$, and there exists a function $g_i(x)\in L^{1}(X)$ such that for all $\lambda\in(0,\infty)^{3}$,
\[
\left|p(x|y)\frac{d}{d\lambda_{i}}\log b_{\lambda}(x|y)\right|\le g_{i}(x),
\]
(regularity for differentiation under the integral sign), we state our goal as obtaining
\begin{equation}
\frac{d}{d\lambda_{i}} H_{c}(p(x|y),b_{\lambda}(x|y))=-\int_{X}p(x|y)\frac{d}{d\lambda_{i}}\log b_{\lambda}(x|y)~\mathrm{d}x,\label{eq:Hc_derivative_statement}
\end{equation}
where
\begin{equation}
\frac{d}{d\lambda_{i}}\log b_{\lambda}(x|y)=\frac{\frac{d}{d\lambda_{i}} b_{\lambda}(x|y)}{b_{\lambda}(x|y)}, \label{eq:new_one_checkpoint_earlier}
\end{equation}
and 
\begin{equation}
\begin{split}
&\frac{d}{d\lambda_{i}} b_{\lambda}(x|y)\\
&\hspace{0.5cm}=\left(\frac{d}{d\lambda_{i}} b_{\lambda}(x,y)\right)\frac{1}{b_{\lambda}(y)}+ b_{\lambda}(x,y)\left(\frac{d}{d\lambda_{i}}\frac{1}{b_{\lambda}(y)}\right).
\end{split}
\label{eq:new_checkpoint_to_later_come_back_to}
\end{equation}
We find that
\[
\begin{split}
\frac{d}{d\lambda_{i}}\frac{1}{b_{\lambda}(y)}&=-\frac{\frac{d}{d\lambda_{i}}b_{\lambda}(y)}{b_{\lambda}(y)^{2}}=-\frac{\int_{X}\frac{d}{d\lambda_{i}}b_{\lambda}(x,y)~\mathrm{d}x}{b_{\lambda}(y)^{2}},
\end{split}
\]
where we again use a dominated convergence assumption, now for $x\mapsto b_{\lambda}(x,y)$, differentiable in $\lambda_{i}$ for almost every $x\in X$, $h_{i}(x)\in L^{1}(X)$ exists such that 
\[
\left|\frac{d}{d\lambda_{i}} b_{\lambda}(x,y)\right|\le h_{i}(x).
\]
Substituted into~\eqref{eq:new_checkpoint_to_later_come_back_to}, we yield
\[
\frac{d}{d\lambda_{i}} b_{\lambda}(x|y)=\frac{\frac{d}{d\lambda_{i}}b_{\lambda}(x,y)}{b_{\lambda}(y)}- b_{\lambda}(x|y)\frac{\int_{X}\frac{d}{d\lambda_{i}}b_{\lambda}(x^{\prime},y)~\mathrm{d}x^{\prime}}{b_{\lambda}(y)}.
\]
Substituting this into~\eqref{eq:new_one_checkpoint_earlier}, yields
\begin{equation}
\begin{split}
\frac{d}{d\lambda_{i}}\log b_{\lambda}(x|y)&=\frac{\frac{d}{d\lambda_{i}}b_{\lambda}(x,y)}{b_{\lambda}(x,y)}-\frac{\int_{X}\frac{d}{d\lambda_{i}}b_{\lambda}(x^{\prime},y)~\mathrm{d}x^{\prime}}{b_{\lambda}(y)} \\
&\hspace{-6mm}=\frac{\frac{d}{d\lambda_{i}}b_{\lambda}(x,y)}{b_{\lambda}(x,y)}-\int_{X}\frac{b_{\lambda}(x^{\prime},y)\frac{d}{d\lambda_{i}}b_{\lambda}(x^{\prime},y)}{b_{\lambda}(x^{\prime},y)b_{\lambda}(y)}~\mathrm{d}x^{\prime}\\
&\hspace{-6mm}=\frac{\frac{d}{d\lambda_{i}}b_{\lambda}(x,y)}{b_{\lambda}(x,y)}-\int_{X}b_{\lambda}(x^{\prime}|y)\frac{\frac{d}{d\lambda_{i}}b_{\lambda}(x^{\prime},y)}{b_{\lambda}(x^{\prime},y)}~\mathrm{d}x^{\prime}.
\end{split}\label{eq:logblambdaderivative1}
\end{equation}
By defining $s_{i}(x,y)=(\frac{d}{d\lambda_{i}}b_{\lambda}(x,y))/b_{\lambda}(x,y)$, and substituting~\eqref{eq:logblambdaderivative1} back into~\eqref{eq:Hc_derivative_statement}, we yield
\[
\begin{split}
&\frac{d}{d\lambda_{i}} H_{c}(p(x|y),b_{\lambda}(x|y))\\
&=-\int_{X}p(x|y)\left(s_{i}(x,y)-\int_{X}b_{\lambda}(x^{\prime}|y)(s_{i}(x^{\prime},y))~\mathrm{d}x^{\prime}\right)\mathrm{d}x \\
&\hspace{1cm}=\int_{X}(b_{\lambda}(x|y)-p(x|y))s_{i}(x,y)~\mathrm{d}x.
\end{split}
\]
This allows us to narrow the search for an explicit derivative to finding an analytical expression for $s_{i}(x,y)$, for every $i\in\{L,P,B\}$.

\subsubsection{Expressing $s_{B}(x,y)$}

For $i=B$, we have
\[
\begin{split}
\frac{d}{d\lambda_{B}}b_{\lambda}(x,y)&=\frac{d}{d\lambda_{B}}e^{\log b_{\lambda}(x,y)}\\
&=e^{\log b_{\lambda}(x,y)}\frac{d}{d\lambda_{B}}\log b_{\lambda}(x,y)\\
&=\frac{1}{\lambda_{B}}b_{\lambda}(x,y)\log b_{\lambda}(x,y),
\end{split}
\]
as $b_\lambda(x, y)=(\int_{X^{k}} \hat{p}_\lambda\left(x_{0:k-1}, x, y\right)\mathrm{d}x_{0:k-1})^{\lambda_B}$. We thus find
\[
s_B(x,y)=\frac{\frac{d}{d\lambda_{B}}b_{\lambda}(x,y)}{b_{\lambda}(x,y)}=\frac{1}{\lambda_{B}}\log b_{\lambda}(x,y).
\]

\subsubsection{Expressing $s_{P}(x,y)$}

For $i=P$, we have
\begin{equation}
\begin{split}
\frac{d}{d\lambda_{P}}b_{\lambda}(x,y)&=e^{\log b_{\lambda}(x,y)}\frac{d}{d\lambda_{P}}\log b_{\lambda}(x,y)\\
&\hspace{-10mm}=\lambda_{B}b_{\lambda}(x,y)\frac{d}{d\lambda_{P}}\log \int_{X^{k}} \hat{p}_\lambda\left(x_{0:k-1}, x, y\right)\mathrm{d}x_{0:k-1},
\end{split} \label{eq:first_eq_lambda_P}
\end{equation}
where, using the dominated convergence assumption for $x_{0:k-1}\mapsto \hat{p}_{\lambda}(x_{0:k-1},x,y)$, we find
\begin{equation}
\begin{split}
&\frac{d}{d\lambda_{P}}\hspace{-0.5mm}\log\hspace{-1.5mm} \int_{X^{k}} \hat{p}_\lambda\left(x_{0:k-1}, x, y\right)\mathrm{d}x_{0:k-1}\\
&\hspace{2cm}=\frac{\int_{X^{k}} \frac{d}{d\lambda_{P}}\hat{p}_\lambda\left(x_{0:k-1}, x, y\right)\mathrm{d}x_{0:k-1}}{\int_{X^{k}} \hat{p}_\lambda\left(x_{0:k-1}^{\prime}, x, y\right)\mathrm{d}x_{0:k-1}^{\prime}},\label{eq:checkpoint_one_for_P}
\end{split}
\end{equation}
and
\[
\begin{split}
&\frac{d}{d\lambda_{P}}\hat{p}_{\lambda}(x_{0:k-1},x,y)=\frac{d}{d\lambda_{P}}e^{\log \hat{p}_{\lambda}(x_{0:k-1},x,y)}\\
&=\hat{p}_{\lambda}(x_{0:k-1},x,y)\frac{d}{d\lambda_{P}}\lambda_{P}\log \hat{p}(y|x_{0:k-1},x)^{\lambda_{L}}\hat{p}(x_{0:k-1},x)\\
&=\frac{1}{\lambda_{P}}\hat{p}_{\lambda}(x_{0:k-1},x,y)\log \hat{p}_{\lambda}(x_{0:k-1},x,y).
\end{split}
\]
Substitution back into~\eqref{eq:checkpoint_one_for_P} then yields
\[
\begin{split}
&\frac{d}{d\lambda_{P}}\log\int_{X^{k}} \hat{p}_\lambda\left(x_{0:k-1}, x, y\right)\mathrm{d}x_{0:k-1}\\
&=\frac{1}{\lambda_{P}}\int_{X^{k}}\hat{p}_{\lambda}(x_{0:k-1}|x,y)\log \hat{p}_{\lambda}(x_{0:k-1},x,y)\mathrm{d}x_{0:k-1},
\end{split}
\]
allowing us to conclude, using~\eqref{eq:first_eq_lambda_P}, that
\[
\begin{split}
&\frac{d}{d\lambda_{P}}b_{\lambda}(x,y)\\
&=\frac{\lambda_{B}}{\lambda_{P}}b_{\lambda}(x,y)\int_{X^{k}}\hat{p}_{\lambda}(x_{0:k-1}|x,y)\\
&\hspace{4cm}\log \hat{p}_{\lambda}(x_{0:k-1},x,y)~\mathrm{d}x_{0:k-1}.
\end{split}
\]
We thus conclude that
\[
\begin{split}
&s_{P}(x,y)=\\
&\hspace{0.6cm} \frac{\lambda_{B}}{\lambda_{P}}\int_{X^{k}}\hat{p}_{\lambda}(x_{0:k-1}|x,y)\log \hat{p}_{\lambda}(x_{0:k-1},x,y)~\mathrm{d}x_{0:k-1}.
\end{split}
\]

\subsubsection{Expressing $s_{L}(x,y)$}

For $i=L$, we have
\begin{equation}
\begin{split}
\frac{d}{d\lambda_{L}}b_{\lambda}(x,y)&=e^{\log b_{\lambda}(x,y)}\frac{d}{d\lambda_{L}}\log b_{\lambda}(x,y)\\
&\hspace{-20mm}=\lambda_{B}b_{\lambda}(x,y)\frac{d}{d\lambda_{L}}\log \int_{X^{k}} \hat{p}_\lambda\left(x_{0:k-1}, x, y\right)\mathrm{d}x_{0:k-1},
\end{split} \label{eq:first_eq_lambda_L}
\end{equation}
where
\begin{equation}
\begin{split}
&\frac{d}{d\lambda_{L}}\hspace{-0.5mm}\log\hspace{-1mm} \int_{X^{k}} \hat{p}_\lambda\left(x_{0:k-1}, x, y\right)\mathrm{d}x_{0:k-1}\\
&\hspace{2cm}=\frac{\int_{X^{k}}\frac{d}{d\lambda_{L}}\hat{p}_{\lambda}(x_{0:k-1},x,y)~\mathrm{d}x_{0:k-1}}{\int_{X^{k}}\hat{p}_{\lambda}(x_{0:k-1}^{\prime},x,y)~\mathrm{d}x_{0:k-1}^{\prime}},\label{eq:checkpoint_one_for_L}
\end{split}
\end{equation}
where
\[
\begin{split}
&\frac{d}{d\lambda_{L}}\hat{p}_{\lambda}(x_{0:k-1},x,y)=\frac{d}{d\lambda_{L}}e^{\log \hat{p}_{\lambda}(x_{0:k-1},x,y)}\\
&=\hat{p}_{\lambda}(x_{0:k-1},x,y)\frac{d}{d\lambda_{L}}\lambda_{P}\log \hat{p}(y|x_{0:k-1},x)^{\lambda_{L}}\hat{p}(x_{0:k-1},x)\\
&=\lambda_{P}\hat{p}_{\lambda}(x_{0:k-1},x,y)\frac{1}{\lambda_{L}}\log \hat{p}(y|x_{0:k-1},x)^{\lambda_{L}},\\
\end{split}
\]
which we substitute back into~\eqref{eq:checkpoint_one_for_L} to get
\[
\begin{split}
&\frac{d}{d\lambda_{L}}\log \int_{X^{k}}\hat{p}_{\lambda}(x_{0:k-1},x,y)~\mathrm{d}x_{0:k-1}\\
&\hspace{0cm}=\frac{\lambda_{P}}{\lambda_{L}}\int_{X^{k}}\hat{p}_{\lambda}(x_{0:k-1}|x,y)\log \hat{p}(y|x_{0:k-1},x)^{\lambda_{L}}~\mathrm{d}x_{0:k-1},
\end{split}
\]
allowing us to conclude, using~\eqref{eq:first_eq_lambda_L}, that
\[
\begin{split}
&\frac{d}{d\lambda_{L}}b_{\lambda}(x,y)\\
&=\frac{\lambda_{P}\lambda_{B}}{\lambda_{L}}b_{\lambda}(x,y)\hspace{-1.5mm}\int_{X^{k}}\hspace{-1mm}\hat{p}_{\lambda}(x_{0:k-1}|x,y)\\
&\hspace{4cm}\log \hat{p}(y|x_{0:k-1},x)^{\lambda_{L}}\mathrm{d}x_{0:k-1}.
\end{split}
\]
We thus have
\[
\begin{split}
&s_{L}(x,y)=\\
&\hspace{4mm}\frac{\lambda_{P}\lambda_{B}}{\lambda_{L}}\hspace{-1.5mm}\int_{X^{k}}\hspace{-1mm}\hat{p}_{\lambda}(x_{0:k-1}|x,y)\log \hat{p}(y|x_{0:k-1},x)^{\lambda_{L}}\mathrm{d}x_{0:k-1}.
\end{split}
\]

\subsubsection{Combining the derivatives}

Using the above expressions $s_{i}(x,y)$, for $i\in\{L,P,B\}$, we state the complete (vector-valued) first NLL derivative as
\[
\begin{split}
&\nabla_{\lambda} H_{c}(p(x|y),b_{\lambda}(x|y))\\
&=\int_{X}(b_{\lambda}(x|y)-p(x|y))\\
&\hspace{1.7cm}\cdot\underset{\hat{p}_{\lambda}(x_{0:k-1}|x,y)}{\mathbb{E}}\begin{bmatrix}
\frac{\lambda_{P}\lambda_{B}}{\lambda_{L}}\log \hat{p}(y|x_{0:k-1},x)^{\lambda_{L}} \\
\frac{\lambda_{B}}{\lambda_{P}}\log \hat{p}_{\lambda}(x_{0:k-1},x,y) \\
\frac{1}{\lambda_{B}}\log b_{\lambda}(x,y)
\end{bmatrix}\mathrm{d}x.
\end{split}
\]


\bibliographystyle{IEEEtran}
\bibliography{references}

\end{document}